\documentclass[aps,prl,reprint,superscriptaddress,longbibliography,raggedbottom,floatfix,nobalancelastpage]{revtex4-2}

\usepackage{amsmath,amssymb,bm}
\usepackage{graphicx}
\usepackage{microtype}
\usepackage{capt-of}
\usepackage{xcolor}
\usepackage[colorlinks=true,citecolor=blue!45!black,
            linkcolor=blue!45!black,urlcolor=blue!45!black]{hyperref}

\newcommand{\vect}[1]{\bm{#1}}
\newcommand{\dd}{\mathop{}\!\mathrm{d}}
\newcommand{\figdir}{figures}

\begin{document}

\title{Granular Rods Fall Faster in Denser Obstacle Fields}

\author{Fumiaki Nakai}
\email[Contact author: ]{fumiaki.nakai@ess.sci.osaka-u.ac.jp}
\author{Hiroaki Katsuragi}
\affiliation{Department of Earth and Space Science, The University of Osaka,
Toyonaka, Osaka 560-0043, Japan}

\date{\today}

\begin{abstract}
How does particle shape affect driven transport through obstacle fields? We simulate a dissipative rod falling under gravity through randomly placed fixed obstacles. For sufficiently slender rods, the mean descent speed before trapping decreases, increases, and then decreases again as obstacle density rises, opening a window of densities in which rods fall faster in denser fields. Scaling arguments based on collision rates and rod geometry explain the three regimes and their crossovers. Particle shape can thus reverse the expectation that crowding slows driven transport.
\end{abstract}

\maketitle

\raggedbottom

\emph{Introduction.---}
The transport of anisotropic objects through crowded environments is a
fundamental problem in statistical physics, relevant to systems across
a wide range of scales, from polymer solutions and liquid crystals to
suspensions, porous media, and granular materials
\cite{Doi1988-te,de-Gennes1971-na,Kasimov2016-ab,Yadav2012-qe}.
For anisotropic objects, rotational and translational motions are hindered to different extents by interactions with their surroundings.
Coupling between these motions can produce transport qualitatively distinct from that of spherical particles
\cite{Frenkel1981,Munk2009-af,Kroger2004-ij,Hofling2008,Petit2026-or,Khalilian2016,Mandal2020-zm,Hofling2008-pi,Moreno2004-xu,Chen2024-cv,Kremer1990-ls}.

The tube picture, in which transverse motion is restricted while motion
along the object's contour remains possible, has provided a foundation
for understanding polymer and rod diffusion at microscopic scales
\cite{de-Gennes1971-na,DoiEdwards1978,Doi1988-te,Kroger2004-ij,Leitmann2016}.
For rods, this confinement also restricts rotation
\cite{DoiEdwards1978,Munk2009-af,Leitmann2016,Moreno2004-qj,Tucker2011}.
The effects of these constraints depend on the underlying dynamics.
For Brownian rods subject to solvent drag and thermal fluctuations,
transverse diffusion is suppressed while axial diffusion is preserved,
so the center-of-mass diffusion coefficient decreases with density
but approaches a finite value
\cite{Munk2009-af}.
In energy-conserving Newtonian systems, by contrast, rotational constraints
can prolong ballistic motion along the rod axis, enhancing translational
diffusion as density increases
\cite{Frenkel1981,Frenkel1983-xy,Magda1986-zg,Hofling2008,Nakai2023,Tucker2010}.
These studies show how constraints on some degrees of freedom
can give rise to diverse diffusion behavior in microscopic systems.

These findings on diffusion of microscopic anisotropic objects motivate us
to ask how translational and rotational constraints affect drift under
external forcing and dissipation, a fundamental question in nonequilibrium
transport in granular and colloidal systems.
Force-driven transport through obstacle fields has been widely studied
\cite{Martin1999-zf,Ghosh2012-xj,Leitmann2013-px,Stoop2018,
Gao2023-vx,Vyas2026-gj,Petit2026-or,Khatri2020-qa},
yet how these constraints shape the density dependence of the mean drift
of anisotropic objects remains unclear.
The drift of an externally driven rod undergoing collisions that constrain
its motion and dissipate energy cannot be inferred from previous findings
on diffusion.

\begin{figure*}[t!]
 \centering
 \includegraphics[width=\textwidth]{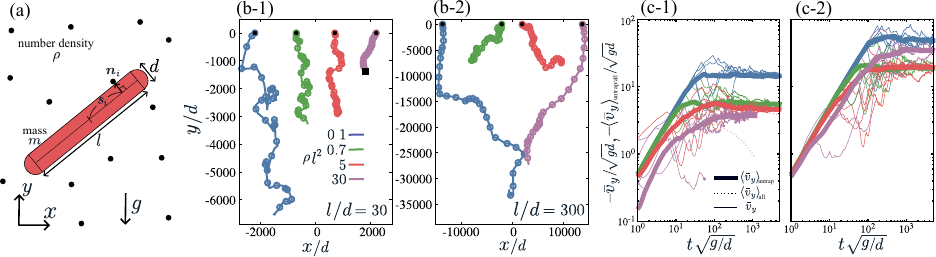}
 \caption{Model and representative dynamics.
(a) A spherocylinder of mass $m$, center-line length $l$, and diameter $d$
falls under gravity $g$ through fixed point obstacles of number density $\rho$.
The coordinate $s_i$ specifies the contact location along the rod axis
for the $i$th obstacle, and $\vect{n}_i$ is the corresponding contact normal.
(b-1),(b-2) Center-of-mass trajectories for $l/d=30$ and $300$, respectively,
over $0\le t\sqrt{g/d}\le600$ at $\rho l^2=0.1,0.7,5,$ and $30$,
shifted horizontally for clarity.
Black circles and squares mark initial and trapping positions, respectively.
The fall distance decreases with density for $l/d=30$,
but first decreases and then increases for $l/d=300$.
See the Supplemental Material~\cite{SupplementalMaterial} for movies of these trajectories.
(c-1),(c-2) Corresponding time-averaged descent speeds,
$-\bar v_y(t)=-t^{-1}\int_0^t v_y(t')\,\mathrm{d}t'$, with the positive $y$ direction pointing upward.
Thin curves show five individual realizations, thick solid curves show
averages over rods that remain untrapped at time $t$, and dashed curves
show averages over all 32 realizations.
For trapped rods, the instantaneous velocity is set to zero after trapping.
The mean descent speed of untrapped rods approaches a plateau at long times;
the plateau value decreases with density for $l/d=30$,
but varies nonmonotonically for $l/d=300$.}
 \label{fig:system}
\end{figure*}

To address this question, we numerically study a simple model: a
dissipative rod falling under gravity through randomly distributed fixed
point obstacles [Fig.~\ref{fig:system}(a)].
We find that, for sufficiently slender rods, the mean fall speed before
trapping decreases, increases, and then decreases again as the obstacle
density rises.
Thus, a rod can fall faster in a denser obstacle field.
We relate this nonmonotonic response to rotational relaxation and
translational velocity persistence, and explain the three regimes and their
crossovers through scaling arguments based on collision rates and rod
geometry.
We attribute the intermediate-density speedup to enhanced velocity persistence associated with hindered rotation, 
and the high-density slowdown to end collisions that limit this persistence.
These mechanisms do not occur for spherical particles.
These results show how the coupling between rotation and translation
arising from particle anisotropy can produce force-driven transport
qualitatively distinct from that of spherical particles.

\emph{Model.---}
We consider a single rod falling under gravity through fixed point obstacles in a
two-dimensional square with periodic boundary conditions
[Fig.~\ref{fig:system}(a)].  The rod is a
spherocylinder of mass $m$, diameter $d$, and center-line length $l$, with
end-to-end length $l+d$.  Assuming a uniform mass density within the rod, its
moment of inertia about its center is
$I=m(dl^3/12+\pi d^2l^2/16+d^3l/4+\pi d^4/32)/(dl+\pi d^2/4)$.
Fixed point obstacles are distributed independently and uniformly with number density $\rho$; using finite-size obstacles is geometrically equivalent to increasing the rod diameter.

We solve the translational and rotational equations of motion of the rod with
a simple contact law consisting of a repulsive elastic force and a dissipative
force proportional to the normal contact velocity.
Let $\vect{v}$, $\vect{u}=(\cos\theta,\sin\theta)$, and
$\omega=\dot\theta$ denote the center-of-mass velocity, unit vector along the rod axis, and
angular velocity of the rod.  For the $i$th point obstacle, let $s_i$ be the
axial coordinate of the closest point on the rod center line, $q_i$ its distance
from the obstacle, and $\vect{n}_i$ the unit vector from the rod toward the
obstacle.  The overlap and normal contact velocity are then
$\delta_i=d/2-q_i$ and
$v_{n,i}=[\vect{v}+\omega\vect{e}_z\times(s_i\vect{u})]\cdot\vect{n}_i$.
The equations of motion are
$m\dot{\vect{v}}=-mg\vect{e}_y+\sum_i\vect{F}_i$ and
$I\dot\omega=\sum_i[(s_i\vect{u})\times\vect{F}_i]_z$, with contact force
$\vect{F}_i=-\Theta(\delta_i)\max(0,k\delta_i+\eta v_{n,i})\vect{n}_i$.
This is the conventional frictionless linear spring--dashpot contact
law used in DEM simulations of spherocylinders \cite{Pournin2005}; the Heaviside step function $\Theta$
and the maximum ensure that it acts only during overlap and remains
repulsive.  We set the nominal
coefficient of restitution to $\epsilon=0.7$, corresponding to
$\eta=-2\ln(\epsilon)\sqrt{km}/\sqrt{\pi^2+[\ln(\epsilon)]^2}$.

We choose units such that $m=d=g=1$ for the simulations,
but retain these quantities explicitly in the expressions for physical clarity. We use $k=5\times10^8$ and a
time step $\Delta t=10^{-6}$.  We integrate the equations using the
velocity-Verlet algorithm in a square simulation box of dimensions
$100l\times100l$.  Runs continue to $T=5000$ for $l\le3000$ and to $T=10000$
for $l=10000$, unless trapping occurs first.
We consider the rod trapped when it remains in contact with at least
one obstacle and satisfies
$[|\vect{v}|^2+\{(l+d)\omega/2\}^2]^{1/2}<10^{-5}$,
with both conditions maintained continuously for $0.5$ time units.
We study
$l/d=10,30,100,300,1000,3000,$ and $10000$ at 25 values spanning
$0.05\le\rho l^2\le500$, retaining $\rho dl\le1$ to exclude regimes dense
enough that a randomly placed rod overlaps more than one obstacle on average.
Each run starts with the rod at the box center, $\theta$ uniformly distributed
on $[0,\pi)$, and no rod--obstacle overlap.  We set $v_y=0$ and
independently sample $v_x$ and $(l+d)\omega/2$ with
uniformly distributed magnitudes between $1.5\sqrt{gd}$ and $3\sqrt{gd}$,
with random signs.
Using this setup, we analyze the rod dynamics over 32 independent
realizations of the obstacle configuration and initial state
for each parameter set.
Hereafter, $\bar v_y(t)$ denotes the time average over $[0,t]$, and
$\langle\bar v_y(t)\rangle$ denotes its average over independent runs.

OpenAI Codex (GPT-5) assisted with implementing and debugging the simulation
and analysis code. The authors determined all physical and analytical choices
and verified the code and results.

\begin{figure*}[t!]
 \centering
 \includegraphics[width=\textwidth]{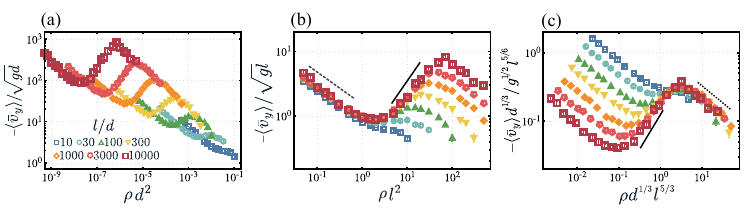}
 \caption{Mean descent speed before trapping. We evaluate
$-\langle\bar v_y\rangle$ at the latest time $t\sqrt{g/d}\le5000$
for which the rod remains untrapped in at least 23 of 32 independent
runs (about 70\%).
 (a) $-\langle\bar v_y\rangle/\sqrt{gd}$ versus $\rho d^2$.  An upturn appears
 for larger $l/d$.  (b) Data scaled
 as $-\langle\bar v_y\rangle/\sqrt{gl}$ versus $\rho l^2$.
 Both the low-density decay and intermediate-density upturn collapse across different $l/d$.
 At higher density, the shorter rods turn down first, followed by
 the longer rods.  The dashed and solid lines are guides with slopes $-1/2$ and $1$,
 respectively.  (c) Data scaled as
 $-\langle\bar v_y\rangle d^{1/3}/(\sqrt{g}l^{5/6})$ versus
 $\rho d^{1/3}l^{5/3}$.  The intermediate- and high-density data collapse.
 The solid and dotted lines are guides with slopes $1$ and $-1/2$,
 respectively.  Error bars show standard errors across the untrapped
 realizations.}
 \label{fig:velocity}
\end{figure*}

\emph{Nonmonotonic drift.---}
To obtain an intuitive picture of the dynamics, we first compare a short rod,
$l/d=30$, with a long rod, $l/d=300$.  Figures~\ref{fig:system}(b-1) and
\ref{fig:system}(b-2) show representative trajectories over
$0\le t\sqrt{g/d}\le600$ at $\rho l^2=0.1,0.7,5,$ and $30$.
For $l/d=30$, the fall distance decreases
monotonically with obstacle density, and trapping occurs at the higher
densities.  For $l/d=300$, by contrast, the fall distance first decreases and
then increases as the density rises.

To test this contrast statistically, we use the 32 independent runs for each parameter set.
For each realization, we define the time-averaged vertical
velocity over $[0,t]$ as
$\bar v_y(t)=t^{-1}\int_0^t v_y(t')\,\dd t'$.  Figures~\ref{fig:system}(c-1)
and \ref{fig:system}(c-2) show five representative realizations as thin
curves, together with averages over all 32 realizations,
$\langle\bar v_y\rangle_{\mathrm{all}}$, and over only those rods that remain
untrapped at time $t$, $\langle\bar v_y\rangle_{\mathrm{untrap}}$.  Starting
from zero vertical velocity, the rods initially accelerate under gravity.
The time-averaged descent speeds then approach plateaus as gravitational work is
balanced, on average, by collisional dissipation.
Short rods are more prone to trapping, whereas long rods rarely become
trapped over the observation period.
Although trapping is not the main focus here, we examine the distance
traveled until trapping in Appendix A.

The run averages confirm the contrasting density dependence of the representative trajectories.
For $l/d=30$
[Fig.~\ref{fig:system}(c-1)], the plateau speed decreases
monotonically with $\rho l^2$, while trapping becomes more frequent.  For
$l/d=300$ [Fig.~\ref{fig:system}(c-2)], the plateau speed instead decreases
and then increases, and trapping is less frequent over the same density
range.

We next quantify how the nonmonotonic descent speed emerges systematically with increasing rod length.
As a measure of the pretrapping descent speed, we evaluate
$-\langle\bar v_y\rangle_{\mathrm{untrap}}$ at the latest time
$t\sqrt{g/d}\le5000$ for which the rod remains untrapped in at least
23 of 32 independent runs (about 70\%).
Long rods rarely become trapped within this time window; for $l/d=10000$,
no trapping is observed in any run at any density studied.
For brevity, we omit the subscript ``untrap'' below.
Figure~\ref{fig:velocity}(a) shows
$-\langle\bar v_y\rangle/\sqrt{gd}$ as a function of $\rho d^2$ for different
$l/d$. For short rods, the descent speed decreases monotonically with
density. A weak upturn first appears around $l/d=30$, and for $l/d\ge100$
the curves clearly show three regimes: the speed decreases, increases,
and then decreases again as density rises.

Replotting the data as $-\langle\bar v_y\rangle/\sqrt{gl}$ versus $\rho l^2$
collapses both the low-density decay and the intermediate-density upturn
[Fig.~\ref{fig:velocity}(b)].
The upturn begins near $\rho l^2\sim1$.
Shorter rods systematically leave the rising
branch at lower $\rho l^2$.  The guides with slopes $-1/2$ and $1$ suggest
the scaling laws $-\langle\bar v_y\rangle\sim\sqrt{g/(\rho l)}$ at low density
and $-\langle\bar v_y\rangle\sim\sqrt{g}\,\rho l^{5/2}$ at intermediate
density.

The intermediate- and high-density data collapse when plotted as
$-\langle\bar v_y\rangle d^{1/3}/(\sqrt{g}l^{5/6})$ versus
$\rho d^{1/3}l^{5/3}$
[Fig.~\ref{fig:velocity}(c)].  The solid and dotted guides, with slopes $1$
and $-1/2$, respectively, correspond to the intermediate-density scaling
above and to $-\langle\bar v_y\rangle\sim\sqrt{g/(\rho d)}$ at high density.
The crossover between these two regimes occurs at
$\rho d^{1/3}l^{5/3}\sim1$.
(The physical origin of these scaling laws is discussed later.)  Thus,
sufficiently long rods exhibit distinct low-, intermediate-, and high-density
regimes, which together produce the nonmonotonic transport.
The nonmonotonic drift persists under moderate tangential friction
[Fig.~\ref{fig:friction} in Appendix B].

\begin{figure}[t]
 \centering
 \includegraphics[width=0.85\columnwidth]{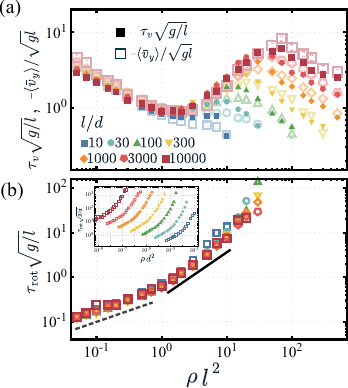}
 \caption{Velocity persistence and rotational relaxation times, defined by the first $1/e$
crossing of the normalized velocity and orientational autocorrelations,
respectively.
(a) $\tau_v\sqrt{g/l}$ (filled symbols) versus $\rho l^2$.
Colors and symbol shapes indicate $l/d$.
The mean descent speed $-\langle\bar v_y\rangle/\sqrt{gl}$ (open symbols)
is shown for comparison and shows a similar density dependence to the
scaled persistence time.
(b) $\tau_{\rm rot}\sqrt{g/l}$ versus $\rho l^2$.
The data suggest a crossover from
$\tau_{\rm rot}\sqrt{g/l}\sim(\rho l^2)^{1/2}$ (dashed)
to $\tau_{\rm rot}\sqrt{g/l}\sim\rho l^2$ (solid)
near $\rho l^2\sim1$.
Inset: $\tau_{\rm rot}\sqrt{g/d}$ versus $\rho d^2$.}
 \label{fig:rotation}
\end{figure}

\emph{Origin of the nonmonotonic drift.---}
To understand the origin of the nonmonotonic speed, we examine translational
velocity persistence and rotational relaxation.
Both correlations use trajectories from the start of each run
until trapping or $t\sqrt{g/d}=5000$, whichever occurs first.
We define $\tau_v$ as the first lag at which the normalized velocity
autocorrelation
$\langle\delta\vect{v}(t_0+\tau)\cdot\delta\vect{v}(t_0)\rangle/
\langle|\delta\vect{v}(t_0)|^2\rangle$
reaches $e^{-1}$.
Here, $t_0$ is a time origin, $\tau$ is the lag time, and $\delta\vect{v}$
is the deviation from the time-averaged velocity within each run,
calculated using only the trajectory before trapping.
The correlation is normalized for each run and then averaged over
independent runs.
Figure~\ref{fig:rotation}(a) compares $\tau_v\sqrt{g/l}$ (filled symbols)
with $-\langle\bar v_y\rangle/\sqrt{gl}$ (open symbols;
the same speed data as in Fig.~\ref{fig:velocity}).
The persistence time closely follows the density dependence of the mean
descent speed, suggesting $-\langle\bar v_y\rangle\sim g\tau_v$.
This suggests that the characteristic descent speed is approximately the velocity gained under gravity over the persistence time $\tau_v$.

For rotational relaxation, we calculate the orientational correlation
$\langle\cos[\theta(t_0+\tau)-\theta(t_0)]\rangle$.
For each lag, the correlation is calculated within each run using only
the trajectory before trapping, then averaged with equal weight over
the available runs.
We define the rotational relaxation time $\tau_{\rm rot}$ as the first
lag at which the averaged correlation reaches $e^{-1}$.
We omit conditions for which the correlation does not reach $e^{-1}$
within the available lag window, $\tau\sqrt{g/d}\le2500$.
The inset of Fig.~\ref{fig:rotation}(b) shows
$\tau_{\rm rot}\sqrt{g/d}$ versus $\rho d^2$.
Rescaling the data as $\tau_{\rm rot}\sqrt{g/l}$ versus $\rho l^2$
collapses the results for different $l/d$ over a broad range
[Fig.~\ref{fig:rotation}(b)].
At higher densities, shorter rods depart from the collapsed curve
at lower $\rho l^2$.
The dashed and solid guides suggest
$\tau_{\rm rot}\sqrt{g/l}\sim(\rho l^2)^{1/2}$ for $\rho l^2\lesssim1$
and $\tau_{\rm rot}\sqrt{g/l}\sim\rho l^2$ for $\rho l^2\gtrsim1$,
respectively.
The crossover near $\rho l^2\simeq1$ marks a stronger slowing of
rotational relaxation with increasing density and coincides with
the upturns in both the mean descent speed and the velocity
persistence time [Fig.~\ref{fig:rotation}(a)].

Before developing the velocity scaling, we identify the basic scales
set by the rod geometry.
Let $v_\ast$ denote a characteristic descent speed.
The rates of side and end collisions scale as
$\tau_{\rm side}^{-1}\sim\rho l v_\ast$ and
$\tau_{\rm edge}^{-1}\sim\rho d v_\ast$, respectively.
A rotation through an angle of order unity sweeps an area of order $l^2$,
containing on average $\rho l^2$ obstacles.
Thus, $\rho l^2\sim1$ marks the onset of geometric constraints on
large reorientations.

In the dilute regime, $\rho l^2\lesssim1$, the rod can rotate substantially
between successive side collisions (see the movies in the Supplemental Material~\cite{SupplementalMaterial}).
Collisions therefore act in different directions, randomizing the
center-of-mass velocity after a few encounters.
The velocity persistence time is then of order $\tau_{\rm side}$, giving
$v_\ast\sim g\tau_{\rm side}$.
Combining this relation with $\tau_{\rm side}^{-1}\sim\rho l v_\ast$
yields $v_\ast\sim\sqrt{g/(\rho l)}$, consistent with the low-density
branch in Fig.~\ref{fig:velocity}(b).
Assuming $l\omega_\ast\sim v_\ast$ for a characteristic angular velocity $\omega_\ast$, 
a large reorientation takes $\tau_{\rm rot}\sim\omega_\ast^{-1}\sim l/v_\ast$.
Thus, $\tau_{\rm rot}\sqrt{g/l}\sim(\rho l^2)^{1/2}$, consistent with
the low-density behavior in Fig.~\ref{fig:rotation}(b).

For $\rho l^2\gtrsim1$, the rod encounters obstacles before it can
substantially reorient.
We extend the tube-like picture developed for diffusion of microscopic rods
\cite{DoiEdwards1978,Frenkel1981,Frenkel1983-xy,Hofling2008,Nakai2023}
to the driven drift of macroscopic rods.
The surrounding obstacles restrict the rod orientation to an angular
range $\delta\theta\sim(\rho l^2)^{-1}$.
After the rod moves a distance of order $l$ along its axis, the tube
is renewed, allowing another angular step of order $\delta\theta$.
Treating successive steps as independent, a large reorientation requires
$N_{\rm renew}(\delta\theta)^2\sim1$, where $N_{\rm renew}$ is the number
of tube renewals.
Thus, $N_{\rm renew}\sim(\delta\theta)^{-2}\sim(\rho l^2)^2$.
Each renewal takes a time of order $l/v_\ast$, so the rotational
relaxation time scales as
$\tau_{\rm rot}\sim(l/v_\ast)N_{\rm renew}\sim\rho^2l^5/v_\ast$.
If a large reorientation also randomizes the translational velocity,
the velocity persistence time is of order $\tau_{\rm rot}$, giving
$v_\ast\sim g\tau_{\rm rot}$.
Combining these relations yields
$v_\ast\sim\sqrt{g}\,\rho l^{5/2}$, which explains the intermediate-density
upturn in Fig.~\ref{fig:velocity}(b).
Equating the low- and intermediate-density speed estimates,
$\sqrt{g/(\rho l)}\sim\sqrt{g}\,\rho l^{5/2}$, gives the crossover
$\rho l^2\sim1$, consistent with the onset of rotational constraints.
Substituting this speed into the expression for $\tau_{\rm rot}$ gives
$\tau_{\rm rot}\sim\rho l^{5/2}/\sqrt{g}$, or equivalently
$\tau_{\rm rot}\sqrt{g/l}\sim\rho l^2$, consistent with the
intermediate-density behavior in Fig.~\ref{fig:rotation}(b).

At higher densities, end collisions can substantially change the
center-of-mass velocity before the rod undergoes a large reorientation.
The velocity persistence time is then of order $\tau_{\rm edge}$, giving
$v_\ast\sim g\tau_{\rm edge}$.
Combining this relation with $\tau_{\rm edge}^{-1}\sim\rho d v_\ast$
yields $v_\ast\sim\sqrt{g/(\rho d)}$, which explains the decreasing
high-density branch in Fig.~\ref{fig:velocity}(c).
The crossover occurs when end collisions limit velocity persistence
before rotational relaxation does.
The condition $\tau_{\rm edge}\sim\tau_{\rm rot}$ gives
$\rho^3dl^5\sim1$, or equivalently
$\rho d^{1/3}l^{5/3}\sim1$, consistent with the crossover observed
in Fig.~\ref{fig:velocity}(c).
Written as $\rho l^2\sim(l/d)^{1/3}$, this crossover shifts to higher
scaled densities as the aspect ratio increases, suggesting unbounded
growth of the descent speed with density in the infinitely thin limit.

Using an empirical expression consistent with the three scaling laws
[Fig.~\ref{fig:regime-map}(a); Appendix C], we summarize the regions
of increasing and decreasing mean descent speed in
Fig.~\ref{fig:regime-map}(b).
This expression yields a nonmonotonic density dependence for
$l/d\gtrsim19$, with a speedup region at intermediate densities
where $\rho l^2$ is of order unity or larger.

Although trapping is not the main focus here, we also analyze the fall
distance until trapping in Appendix A.
The mean distance approximately follows
$-\langle y_{\rm trap}\rangle\sim(\rho^3d^3l^2)^{-1}$,
which we briefly explain using the geometry of three-contact trapping.

\emph{Conclusion.---}
We have shown that long dissipative rods can fall faster in denser
obstacle fields. The mean descent speed before trapping exhibits
three regimes: it decreases, increases, and then decreases again
as obstacle density rises.
Scaling arguments based on side collisions, rotational constraints,
and end collisions explain these regimes and their crossovers.
These results show that constraints on some degrees of freedom arising
from particle shape can produce nonequilibrium transport qualitatively
distinct from that of spherical particles.

\begin{acknowledgments}
\emph{Acknowledgments.---}
This work was supported by JSPS KAKENHI Grant Numbers JP24KJ0156 and JP25K17359.  The
computation in this work was performed using the facilities of the
Supercomputer Center, the Institute for Solid State Physics, the University
of Tokyo (ISSPkyodo-SC-2026-Ba-0021).
\end{acknowledgments}

\emph{Data availability.---}
The data and code supporting the results of this study are
available on Zenodo~\cite{NakaiZenodo2026}.

\bibliography{references}

\clearpage
\appendix
\onecolumngrid
\begin{center}
\textbf{End Matter}
\end{center}

\noindent\begin{minipage}[t]{0.48\textwidth}
\vspace{0pt}
\refstepcounter{section}\label{app:trapping}
\emph{Appendix A: Trapping---}Figure~\ref{fig:trapping}(a) shows the survival probability
$P(-y_{\rm trap}>-y)$, calculated from 32 independent trajectories
per parameter set, where $-y$ is the downward fall distance.
Colors and symbols indicate $l/d$ and $\rho l^2$, respectively.
At fixed $\rho l^2$, shorter rods become trapped over shorter distances.
Increasing obstacle density at fixed rod length also shortens the
distance to trapping.
Rescaling the distance by $\rho^3l^2d^3$ approximately collapses
the survival curves.
For conditions in which at least three of the 32 rods become trapped
by $t\sqrt{g/d}=5000$, we obtain the mean trapping distance from
exponential fits to the survival curves.
Plotting $-\langle y_{\rm trap}\rangle l^2/d^3$ versus $\rho d^2$
approximately collapses the data onto a single curve with slope $-3$,
giving $-\langle y_{\rm trap}\rangle\sim(\rho^3d^3l^2)^{-1}$
[Fig.~\ref{fig:trapping}(b)].

This scaling can be explained by a geometric estimate.
Three contacts are expected to be needed to arrest the two translational
and one rotational degrees of freedom.
Side contacts alone cannot prevent axial sliding, so for $l\gg d$,
we expect trapping to be dominated by two side contacts and one end contact.
In the tube-like regime, the probability of two side contacts scales as
$(\rho dl)^2$, while an end encounters a third obstacle at a rate of order
$\rho d$ per unit fall distance.
The trapping rate per unit distance is therefore of order
$\rho^3d^3l^2$, whose inverse gives the scaling of the mean trapping distance.

\par\medskip
\refstepcounter{section}
\emph{Appendix B: Robustness to contact friction---}The main text considers frictionless contacts.
To test the robustness of the nonmonotonic descent speed to friction,
we introduce a tangential spring--dashpot force subject to the Coulomb
criterion $|F_t|\le\mu F_n$ \cite{Cundall1979}, where $F_n$ is the
magnitude of the normal force and $\mu$ is the friction coefficient.
For each contact, we integrate the tangential relative velocity $v_t$
to obtain the displacement $\xi_t$ and calculate the trial force
$F_t^{\rm trial}=-k_t\xi_t-\eta_t v_t$.
We use $F_t=F_t^{\rm trial}$ when $|F_t^{\rm trial}|\le\mu F_n$,
and $F_t=\mu F_n\operatorname{sgn}(F_t^{\rm trial})$ otherwise.
The total contact force is $\vect{F}=-F_n\vect{n}+F_t\vect{t}$,
where $\vect{t}$ is the unit contact tangent.
The tangential velocity and torque are evaluated at the rod surface,
with lever arm $s_i\vect{u}+(d/2)\vect{n}_i$.
The accumulated tangential displacement is reprojected onto the
instantaneous contact tangent as the contact normal changes.
We use $k_t=2k/7$ and $\eta_t=2\eta/7$, with all other parameters
unchanged from the main text.
For $\mu=0$, this recovers the frictionless contact law.

Figure~\ref{fig:friction} shows the mean descent speed before trapping
for $\mu=0.1$, $0.3$, and $1$.
The nonmonotonic density dependence remains clear for $\mu=0.1$ and $0.3$,
showing that the speedup is not limited to frictionless contacts.
For $\mu=1$, we attribute the weaker speedup to friction slowing axial motion.
\end{minipage}\hfill
\begin{minipage}[t]{0.48\textwidth}
\vspace{0pt}
\centering
 \includegraphics[width=0.72\linewidth]{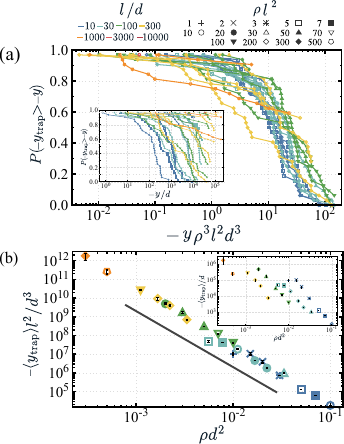}
 \captionof{figure}{Distance to trapping, with $y=0$ at the initial position.
(a) Survival probability $P(-y_{\rm trap}>-y)$ versus
$-y\rho^3l^2d^3$. Colors and symbols indicate $l/d$ and $\rho l^2$,
respectively. The rescaling approximately collapses the curves.
Inset: the same data versus $-y/d$.
(b) Scaled mean trapping distance
$-\langle y_{\rm trap}\rangle l^2/d^3$ versus $\rho d^2$,
obtained from exponential fits to the survival curves.
The solid line has slope $-3$.
Inset: $-\langle y_{\rm trap}\rangle/d$ versus $\rho d^2$.
Error bars show standard errors from the fits.}
 \label{fig:trapping}

\vspace{1em}
\centering
 \includegraphics[width=0.56\linewidth]{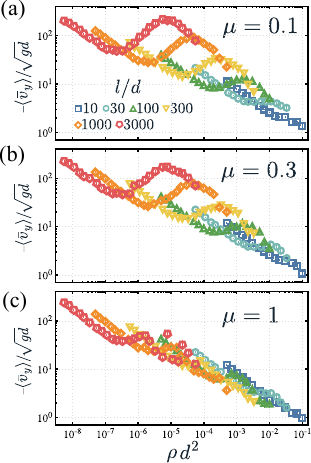}
 \captionof{figure}{Effect of contact friction on the mean descent speed before trapping
for (a) $\mu=0.1$, (b) $\mu=0.3$, and (c) $\mu=1$.
Colors and symbol shapes indicate $l/d$.
The nonmonotonic density dependence remains clear for $\mu=0.1$ and $0.3$,
but is weakened for $\mu=1$.}
 \label{fig:friction}
\end{minipage}

\clearpage
\onecolumngrid
\noindent\begin{minipage}[t]{0.48\textwidth}
\vspace{0pt}
\refstepcounter{section}\label{app:regime-map}
\emph{Appendix C: Empirical interpolation and regime map---}To connect the three asymptotic regimes, we use the empirical expression
$-\langle\bar v_y\rangle/\sqrt{gl}
=[c_1(\rho l^2)^{-1/2}+c_2\rho l^2]/
[1+(c_2/c_3)(\rho l^2)^{3/2}\sqrt{d/l}]$.
The numerator describes the low- and intermediate-density behavior; the denominator accounts for finite rod thickness.
The expression gives
$-\langle\bar v_y\rangle\simeq c_1\sqrt{g/(\rho l)}$,
$c_2\sqrt{g}\,\rho l^{5/2}$, and $c_3\sqrt{g/(\rho d)}$
in the low-, intermediate-, and high-density regimes, respectively,
reproducing the three scaling laws in Fig.~\ref{fig:velocity}.
A common fit to all 153 parameter sets gives
$c_1=0.796$, $c_2=0.213$, and $c_3=0.728$.
With these coefficients, the expression agrees reasonably with
the simulation data [Fig.~\ref{fig:regime-map}(a)].

Figure~\ref{fig:regime-map}(b) shows the speed calculated from the
empirical expression as a color map, with black boundaries separating
the regions of increasing and decreasing speed.
White denotes the excluded region $\rho ld>1$.
Examining the sign of the partial derivative with respect to density
gives the condition $l/d>(4c_1/c_3)^2\simeq19.2$ for a speedup region
to exist.
The speed increases with density between the two boundaries and
decreases outside them.
Using the fit coefficients, the boundaries merge at
$\rho l^2=(2c_1/c_2)^{2/3}\simeq3.82$ and $l/d\simeq19.2$.
For sufficiently long rods, the speedup region is approximately
$1.52<\rho l^2<3.60(l/d)^{1/3}$.

\end{minipage}\hfill
\begin{minipage}[t]{0.48\textwidth}
\vspace{0pt}
 \centering
 \includegraphics[width=0.85\linewidth]{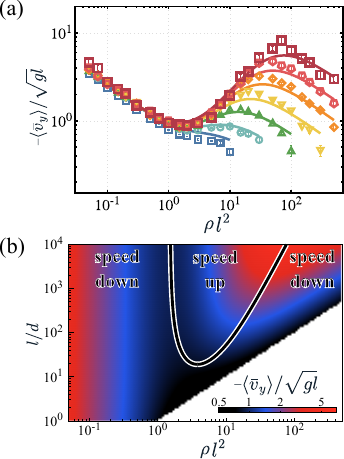}
 \captionof{figure}{Empirical interpolation of the descent speed and its density-dependent
increase and decrease regions.
(a) Mean descent speed before trapping
$-\langle\bar v_y\rangle/\sqrt{gl}$ versus $\rho l^2$.
Symbols show the same simulation data as in Fig.~\ref{fig:velocity}.
Curves of matching colors show the empirical expression with coefficients
shared across all rod lengths and agree reasonably with the data.
(b) Descent speed obtained from the empirical expression in the
$(\rho l^2,l/d)$ plane. Color represents $-\langle\bar v_y\rangle/\sqrt{gl}$
on a logarithmic scale.
Black boundaries separate the regimes of decreasing, increasing, and
decreasing speed as density rises, with nonmonotonic behavior emerging
for $l/d\gtrsim19$.
White denotes the excluded region $\rho ld>1$.}
 \label{fig:regime-map}
\end{minipage}

\end{document}